# Content Based Image Retrieval Using Mobile Agents and Steganography

Sabu .M Thampi
*Assistant Professor*
*Department of Computer Sc. & Engg.*
*L.B.S College of Engineering,*
*Kasaragod, Kerala-671542*
*smtlbs@yahoo.co.in*

Dr. K. Chandra Sekaran
*Professor & Head*
*Department of Computer Engineering*
*National Institute of Technology Karnataka,*
*Surathkal - 574157*
*kch@nitk.ac.in*

## Abstract

*In this paper we present an image retrieval system based on Gabor texture features, steganography, and mobile agents.. By employing the information hiding technique, the image attributes can be hidden in an image without degrading the image quality. Thus the image retrieval process becomes simple. Java based mobile agents manage the query phase of the system. Based on the simulation results, the proposed system not only shows the efficiency in hiding the attributes but also provides other advantages such as: (1) fast transmission of the retrieval image to the receiver, (2) searching made easy.*

## 1. Introduction

The recent tremendous growth in computer technology has also brought a substantial increase in the storage of digital imagery. Examples of applications can be found in every day life, from museums for archiving images or manuscripts, to medicine where millions of images are generated by radiologists every year. Storage of such image data is relatively straightforward, but an accessing and searching image database is intrinsically harder than their textual counterparts. The goal of Content-Based Image Retrieval (CBIR) systems is to operate on collections of images and, in response to visual queries, extract relevant image. The application potential of CBIR for fast and effective image retrieval is enormous, expanding the use of computer technology to a management tool.

The main problem considered in a content-based image retrieval system is how to effectively and efficiently find an image from a large number of images in an image database.

Strategies proposed for image retrieval systems basically approach content-based algorithms by utilizing the similarity between the images in the database and the query image. Similarity of two images can be defined as the distance of their features. In most of the systems developed so far, after the retrieval, only the image data and not the attribute information is delivered to the user terminal. Unfortunately, since searching and browsing may not be the end tasks in many applications, the attribute information is useful for further implementations. Therefore, it is more practical, if the attribute information is transmitted along with the image data. However, transmitting the attributes to the end user requires additional transmission cost, data management and storage space. To overcome this problem, we utilize the concept of information hiding (steganography) to attach the image attributes in the image without increasing the image data size. Thus, this new steganographic approach increases the functional capability of retrieval systems, which allow an end user to perform further searches by directly using the embedded attributes in the retrieved image.

The proposed framework for image retrieval from image databases utilises the newest technology of mobile agents, Gabor texture features, and information hiding techniques (steganography). The speed of image retrieval is increased to some extent with the addition of mobile agents.

The remaining sections of the paper are organised as follows. Section 2 provides an overview of mobile agents. Section 3 describes CBIR algorithm using Gabore Texture features. DCT information hiding process is explained in section 4. Section 5 presents the proposed image retrieval system. Section 6 demonstrates the proposed system. Results and discussion are presented in section 7. Section 8 concludes the paper.

## 2. Mobile agents

Mobile agents are processes dispatched from a source computer to accomplish a specified task. Mobility increases the functionality of the mobile agent. A mobile agent consists of the program code and the program execution state. Initially a mobile agent resides on a computer called the home machine. The agent is then dispatched to execute on a remote computer called a mobile agent host. When a mobile agent is dispatched the entire code of the mobile agent and the execution state of the mobile agent is transferred to the host. The host provides a suitable execution environment for the mobile agent to execute. Another feature of mobile agent is that it can be cloned to execute on several hosts. Upon completion, the mobile agent delivers the results to the sending client or to another server.

Aglet Technology is a framework for programming mobile network agents in Java developed by the IBM™ Japan research group. The IBM's mobile agent is called 'Aglet', is a lightweight Java object. An aglet can be dispatched to any remote host that supports the Java Virtual Machine. This requires from the remote host to pre-install Tahiti, a tiny aglet server program implemented in Java and provided by the Aglet Framework. To allow aglets to be fired from within applets, the IBM Aglet team provided the so-called "FijiApplet", an abstract applet class that is part of a Java package called "Fiji Kit". FijiApplet maintains some kind of an aglet context. From within this context, aglets can be created, dispatched from and retracted back to the FijiApplet.

## 3. CBIR using Gabor texture features

This section describes an image retrieval technique based on Gabor texture features.. In contrast to colour, which is a point, feature texture is a region feature. It relates mostly to a specific spatially repetitive structure of surfaces formed by one or more primary elements. Generally the repetition may involve local variations of scale orientation or other geometric and optical features of the elements.

Gabor filter (or Gabor wavelet) is widely adopted to extract texture features from the images for image retrieval and has been shown to be very efficient. Basically, Gabor filters are a group of wavelets, with each wavelet capturing energy at a specific frequency and a specific direction. Expanding a signal using this basis provides a localized frequency description, therefore capturing local features/energy of the signal. Texture features can then be extracted from this group of energy distributions. The scale (frequency) and orientation tunable property of Gabor filter makes it especially useful for texture analysis. A rotation normalization method that achieves rotation invariance by a circular shift of the feature elements so that all images have the same dominant direction is proposed here.

### 3.1 Texture representation

After applying Gabor filters on the image with different orientation at different scale, we obtain an array of magnitudes:

$$E(m, n) = \sum_x \sum_y | G_{mn}(x,y) |,$$
$$m = 0, 1 \ldots M-1; n = 0, 1 \ldots N-1$$

These magnitudes represent the energy content at different scale and orientation of the image.

The main purpose of texture-based retrieval is to find images or regions with similar texture. It is assumed that we are interested in images or regions that have homogenous texture, therefore the following mean $\mu_{mn}$ and standard deviation $\boldsymbol{\sigma}_{mn}$ of the magnitude of the transformed coefficients are used to represent the homogenous texture feature of the region:

$$\mu_{m,n} = E(m,n) / P \times Q$$

$$\sigma_{mn} = \frac{\sqrt{\sum_x \sum_y (| G_{mn}(x,y) | - \mu_{m,n})^2}}{P \times Q}$$

A feature vector f (texture representation) is created using $\mu_{m,n}$ and $\boldsymbol{\sigma}_{mn}$ as the feature components.

Five scales and 6 orientations are used in common implementation and the feature vector is given by:

$$\mathbf{f} = (\mu_{00}, \sigma_{00}, \mu_{01}, \sigma_{01} \ldots \mu_{45}, \sigma_{45}).$$

### 3.2 Rotation invariant similarity measurement

The texture similarity measurement of a query image $Q$ and a target image $T$ in the database is defined by:

$$D(Q, T) = . \sum_m \sum_n . d_{mn}(Q,T)$$

Where

$$d_{mn} = \sqrt{(\mu^Q_{mn} - \mu^T_{mn})^2 + (\sigma^Q_{mn} - \sigma^T_{mn})^2}$$

Since this similarity measurement is not rotation invariant, similar texture images with different direction may be missed out from the retrieval or get a low rank.
A simple circular shift on the feature map is sufficient to solve the rotation variant problem associated with Gabor texture features. Specifically, we calculate total energy for each orientation. The orientation with the highest total energy is called the dominant orientation/direction. We then move the feature elements in the dominant direction to be the first elements in f. The other elements are circularly shifted accordingly. For example, if the original feature vector is *"abcde*f" and *"*c" is at the dominant direction, then the normalized feature vector will be *"cdefa*b". This normalization method is based on the assumption that to compare similarity between two images/textures they should be rotated so that their dominant directions are the same.

## 4. DCT based information hiding process

Transform coding is simply the compression of images in the frequency domain. The Discrete Cosine Transform (DCT) is an example of transform coding. JPEG transforms the information of color domain into frequency domain by applying Discrete Cosine Transform (DCT). The image is divided into blocks with 8X8 pixels, which is transformed into frequency domain. Each block of an image is represented by 64 components, which are called DCT coefficients. The global and important information of an image block is represented in lower DCT coefficients, while the detailed information is represented in upper coefficients. The compression of an image is achieved by omitting the upper coefficients.

Steganography is the art and science of communicating in a way, which hides the existence of the communication. Transform Domain based information hiding methods utilize an algorithm such as the Discrete Cosine Transformation (DCT) to hide information in significant areas of the image. The image attributes are stored within the cover image.

The algorithms for embedding and extracting one bit of information with DCT are given below: The cover medium is an image.

**Embedding process**

- Compute the DCT coefficients for each 8x8 block
- Quantize the DCT coefficients by standard JPEG quantization table
- Modify the coefficients according to the bit to hide
    - If bit=1, all coefficients are modified to odd numbers
    - If bit=0, all coefficients are modified to even numbers
    - All coefficients quantized to 0 are remain intact
- Inverse quantization
- Inverse DCT

**Extracting process**

- Compute the DCT coefficients for each 8x8 block
- Quantize the DCT coefficients by standard JPEG quantization table
- Count the numbers of coefficients quantized to odd and even
    - If odd coefficients are more, then bit=1
    - If even coefficients are more, then bit=0

The embedding process and extracting process is illustrated in a flowchart (figure 1).

## 5. Proposed Image Retrieval System

The proposed image retrieval system is designed based on the details given above. As illustrated in figure 2 the system consists of attributes generation and embedding phase, and query processing phase. The attribute generation and embedding phase serves as a preprocessing operation for the image retrieval system. Its functions are to prepare, arrange, store and manage information in the database. This phase also performs the general operations such as texture feature extraction and feature embedding. Texture features are extracted from the image by the procedure described in section 3. The rotation variant problem associated with Gabor texture features is managed by a circular shift The image attributes are embedded into the image by an algorithm described in section 4.

On the other hand, the query phase serves as the user interface. In this phase the user enters a query. The CBIR service returns a list of image descriptors of images matching the query in the order of similarity. The images themselves must be retrieved from the providers. Since the attributes are embedded in the image data, the attribute extraction is required to decode the embedded attributes before performing the query process. Only the images, which have similar characteristics as query image, are retrieved and transmitted to the end user. The Image-applet is responsible for forming a graphical client database interface that user can employ to enter image requests. The Image-applet is an extension of the abstract FijiApplet class. The Image-Aglet is created within the context of the Image-applet and is responsible for carrying the users request to the Broker server through the web server, executing it, and returning the results back to the Image-applet context. The Image-Aglet is a Java-based extension of the Aglet class. Other agents associated with the system are image-index agent, image-search agent, parked agent and messenger agent.

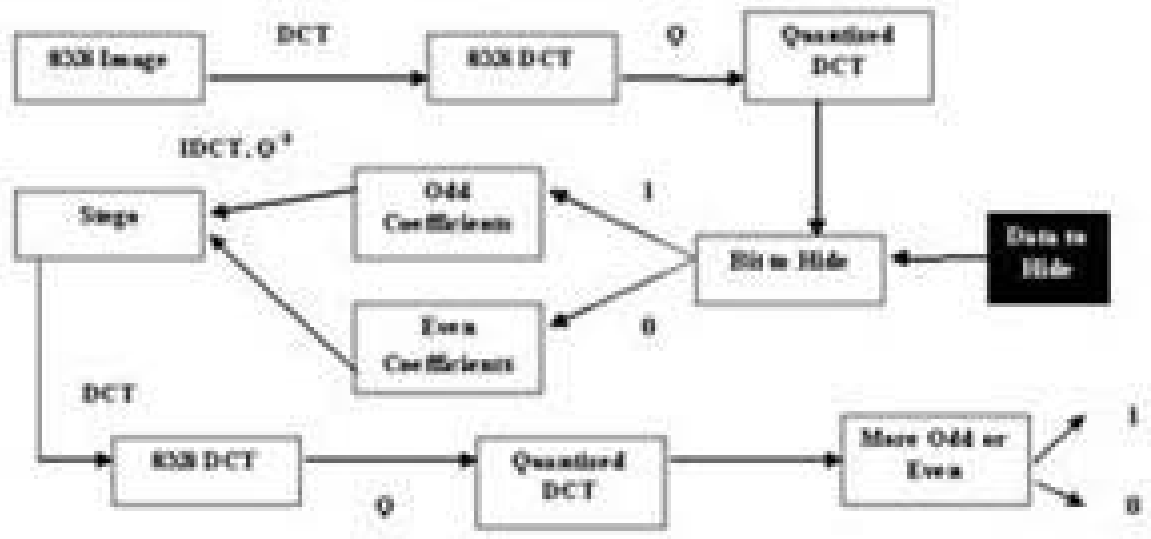

**Figure 1. Embedding and Extracting Process**

## 6. Demonstrating the Image Retrieval System

To demonstrate first download at the Client host the html page containing the image-applet and the Image-Aglet. The user selects or inputs a query image through the Image-applet. The image attributes are computed and embedded into the image. Two Image-Aglets are fired from the Image-applet. The first one is called *parked Image-Aglet*. The parked Image-Aglet carries the following message directions:

- The address of the URL where the broker server is located.
- Query to be executed.
- The appropriate certificates for the aglet to be trusted at the broker server.

The role of the parked agent is to camp at the broker server's agent context, submit the client's request, load the appropriate drivers and collect the answer for the query. The second Image-Aglet is called the messenger aglet. The messenger aglet is responsible for carrying the result back to the Image-Aglet (see figure 2). Any subsequent requests are transmitted via the messenger aglet to the parked Image-Aglet.

As soon as the Broker server receives user request it sends image-index agents to the image providers' servers. Multiple copies of index agents may be created to visit servers of various image providers. The index agents transport the attribute extraction and collection algorithms. Index agents may compute and collect indexes of multiple images archives, which are sent back to the broker server. The parked agents collect the indexes.

The client receives a list of image descriptors of images matching the query in the order of similarity. Each image descriptor consists at least of a thumbnail, the image identifier of that image (unique within the domain of the image provider), a measure of "similarity" to the original query image, and the URL of the Provider's agent server from which the image can be retrieved. The image themselves must be retrieved from the providers.

User can send requests to retrieve the images through the applet after viewing the thumbnails of images. The messenger aglet carries these requests to the broker server. The broker server dispatches a new type of mobile agents called search agents. Multiple copies of search agents are created to visit Providers' sites. Search agents are allowed to retrieve full images from image databases of various image providers.

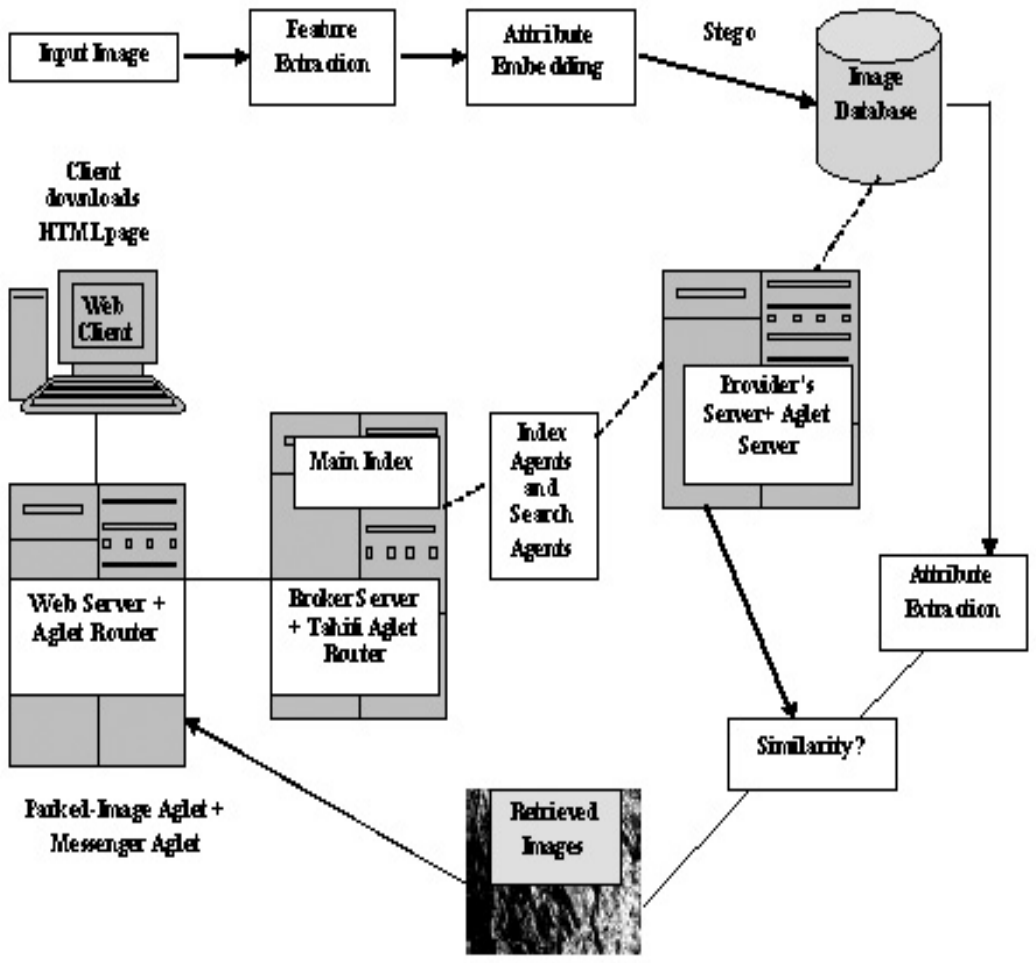

**Figure 2. The Proposed Image Retrieval System**

## 7. Results and Discussion

Accuracy, stability and speed are considered as features of an efficient content-based retrieval scheme. As far as accuracy of the retrieved images is concerned the system works well for images with homogeneous texture features. But the performance is seriously damaged for higher noise (i.e. noise power higher than 50). The image retrieval rate is higher than the applet based technique due to the application of mobile agent technology.

Signal-to-noise measures are estimates of the quality of a reconstructed image compared with an original image. The invisibility of the embedded attributes is evaluated by measuring the PSNR. PSNR is used to measure the difference between two images.

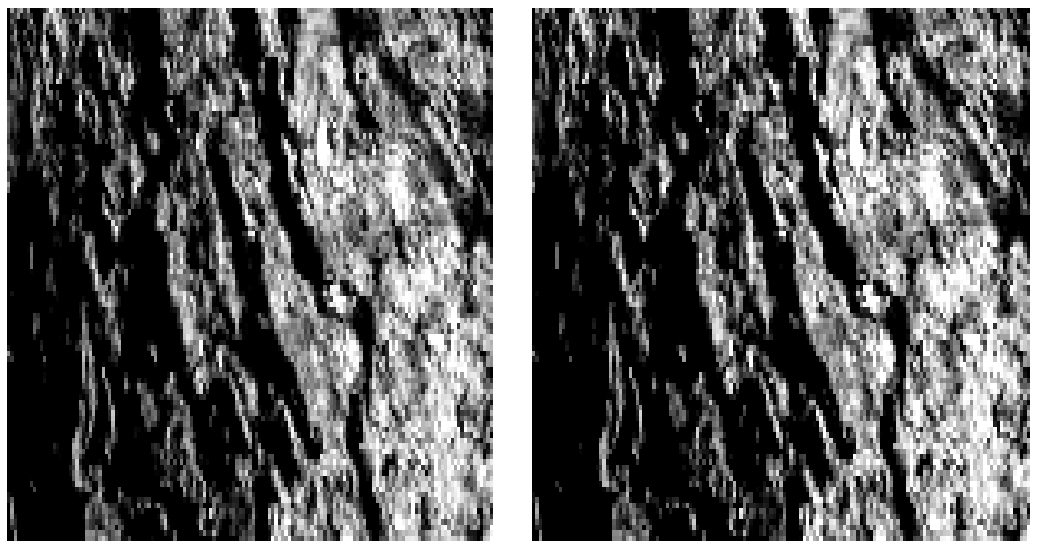

**Figure 3. Original image (PSNR=36.78 dB) and JPEG compressed image with 2000 bits embedded (PSNR=35.69 dB)**

Even though the attributes are embedded into the image (see figure 3) the PSNR of the image changes by only 1.09 dB. In figure 4 the effects on the image fidelity against several embedded bits are compared. As the results, even though we increase the amount of embedded bits by 10Kbits, the PSNR changes slightly. The histograms of the original as well as the stego are plotted in figure 5 and 6.

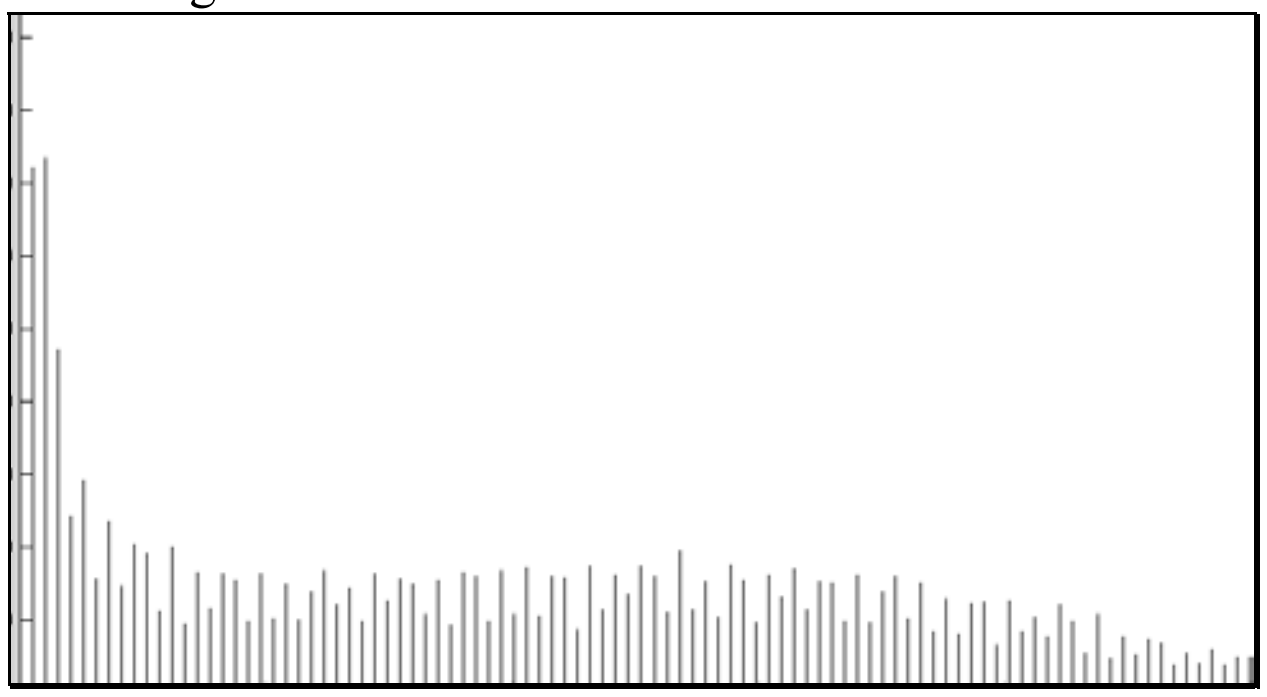

**Figure 5. Histogram of original rock image**

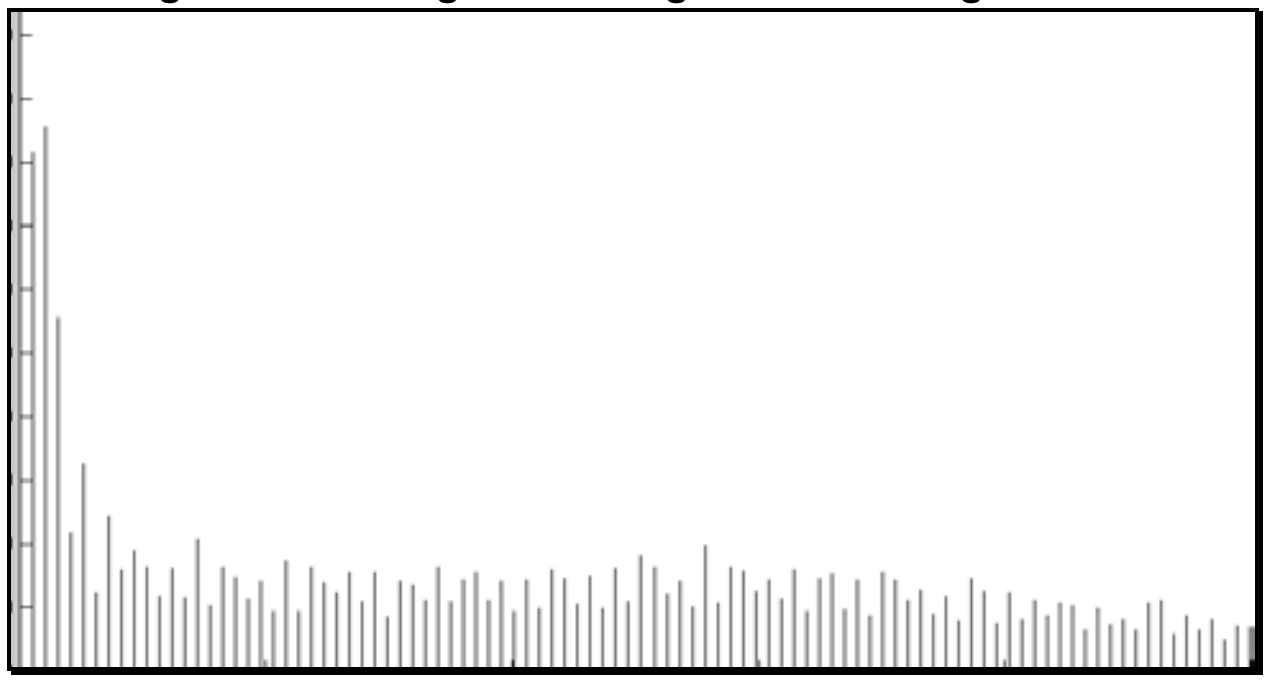

**Figure 6. Histogram of rock image with attributes embedded**

A comparison of the total time required by a web client to access and query image databases between the traditional applet-based and Image-Aglet methodologies are plotted in a graph (see figure 7). For each methodology the client is accessing the Web server via a 64 kbps dial-up connection to an ISP (BSNL). Several tests were conducted with different queries. From the graph we can conclude that Image-Aglet method is the most efficient methodology for all cases of client connectivity.

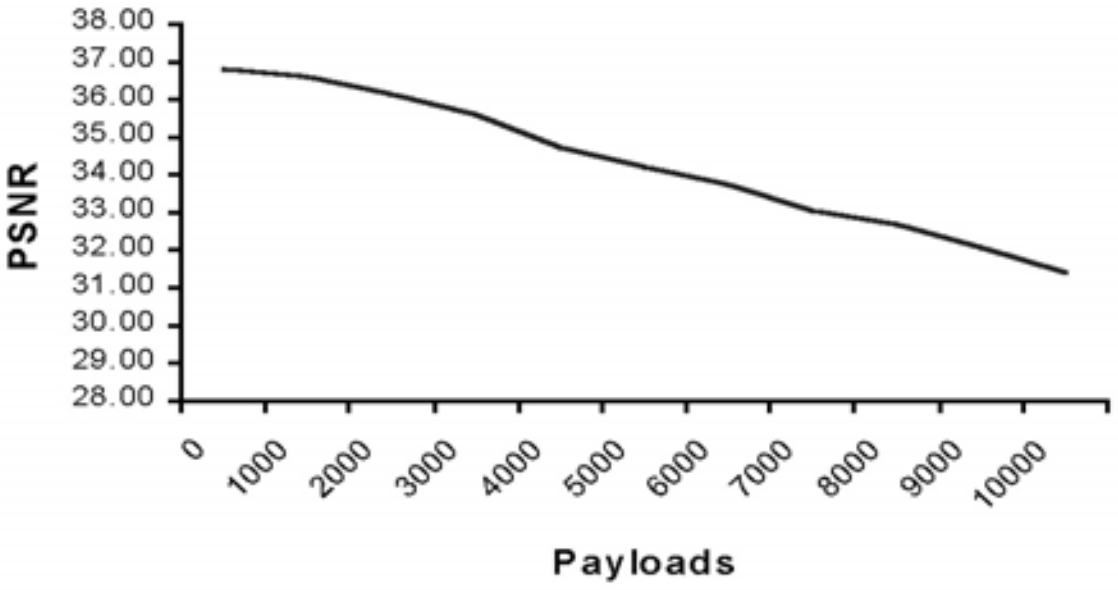

**Figure 4. Payloads vs. PSNR graph for texture image**

Figure 8 shows our preliminary results on image retrieval using the proposed system. The top left image is the query image and the other images are retrieved images from the image database. The retrieved images are ranked in decreasing order based on the similarity of their Gabor texture features to those of the query image.

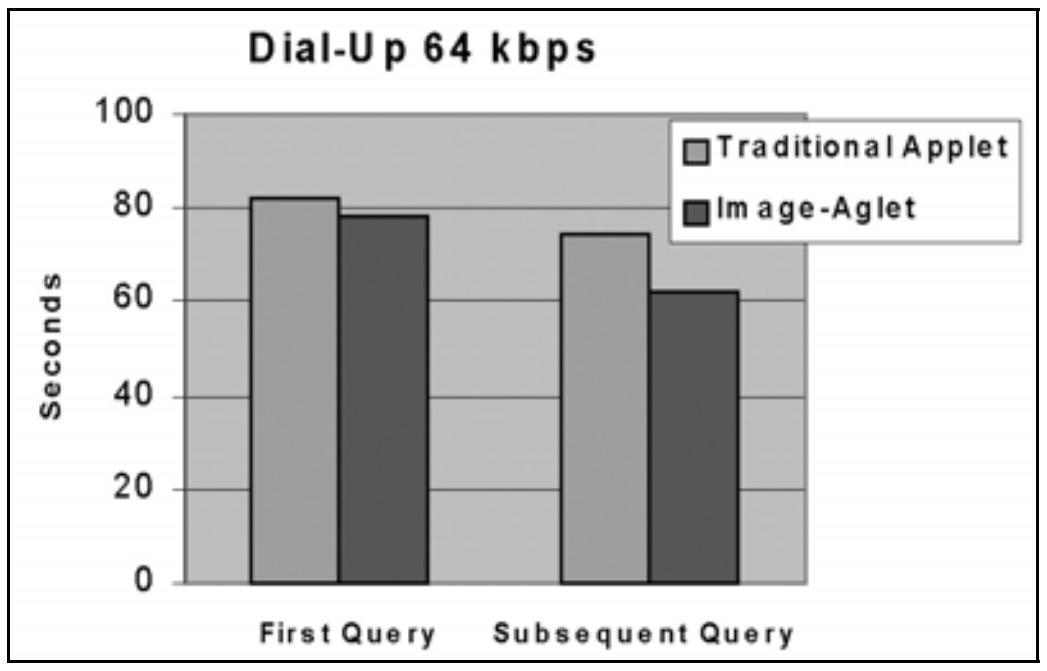

**Figure 7. Mean times for 64 kbps client connectivity**

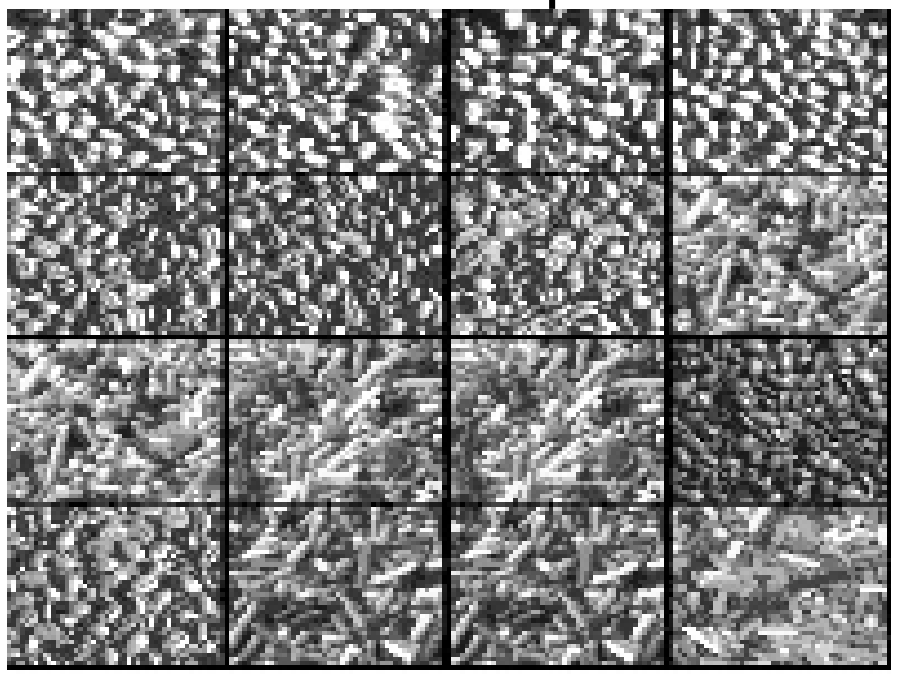

**Figure 8. Preliminary results on image retrieval**

Evaluation of retrieval performance is a crucial problem in content-based retrieval. The common evaluation measures used in CBIR are Precision graph and Recall graph. Precision is the ability to retrieve top-ranked images that are mostly relevant. Recall is the ability of the search to find all of the relevant items in the corpus.

The precision graph gives a better indication of what might be a good number of relevant images for a given query. The recall graph looks more positive than other graphical methods, especially when a few relevant images are retrieved late.

The retrieval performance of the proposed system was evaluated and the results are shown in figures 9 and 10.

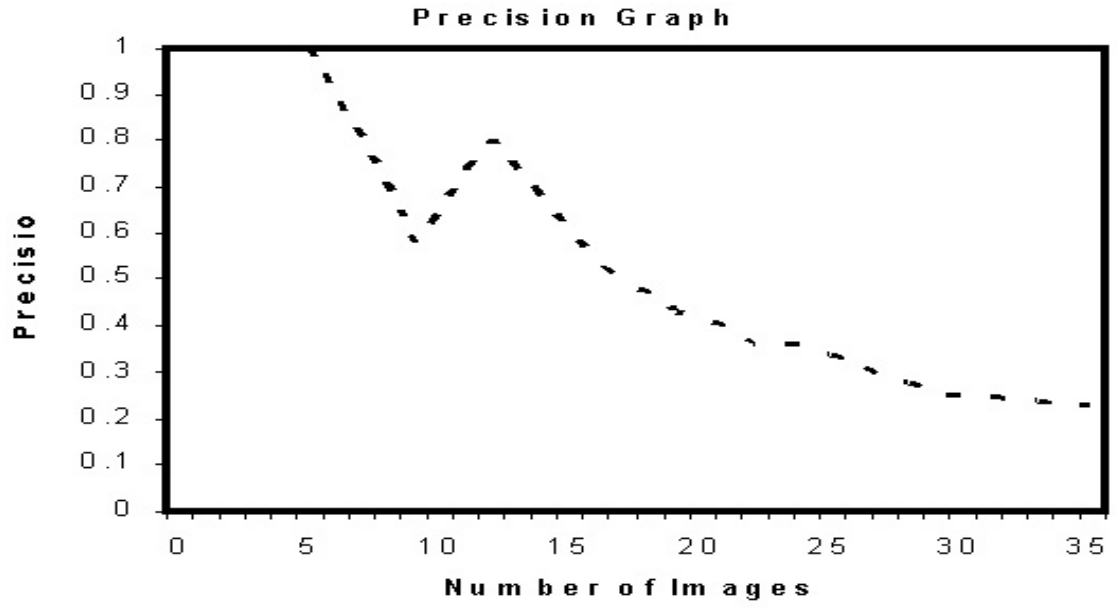

**Figure 9. Precision Graph**

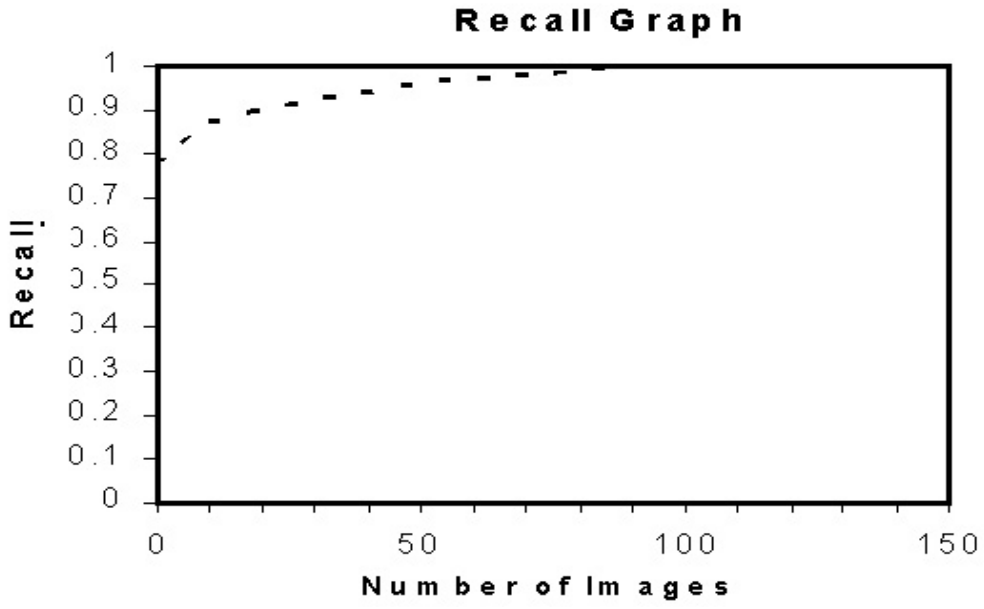

**Figure 1. Recall Graph**

## 8. Conclusions and Future Work

In this paper a new approach for image retrieval is proposed. Preliminary retrieval results have been shown and examined. While the proposed steganographic technique yields attributes that can be invisibly hidden in the image, it also allows reliable and fast decoding of the embedded attributes. The application of mobile agents provides faster data transmission. In case of feature extraction stage global texture features are extracted from the entire image. Hence in our future work texture segmentation will be incorporated to facilitate region-based retrieval. Other techniques which minimize the noise can be applied to achieve better information hiding in images. The proposed system has a centralized indexing system in which indexing agents compute and collect indexes of multiple image archives. These indexes are sent back and merged to the broker's main index. An alternative system can be developed with the intention that the agents may offer CBR services directly from the image provider's host, which completely eliminates the need to transfer indexes.

## 9. References

[1] B. S. Manjunath and W. Y. Ma. "Texture features for browsing and retrieval of large image data," IEEE Transactions on Pattern Analysis and Machine Intelligence, (Special Issue on Digital Libraries), Vol. 18 (8), August 1996, pp. 837-842.

[2] Belmon, S. G., and Yee, B. S, "Mobile agents and intellectual property protection," In Rothermel and Hohl [9], pp. 172–182.

[3] RothermeL, K., and Popescu-Zeletin, R, Eds. *Mobile Agents (MA '97*), vol. 1219 of *Lecture Notes in Computer Scienc*e. Springer Verlag, Berlin Heidelberg, 1997.

[4] Smith,J.R.,and Chang, S.-F, "VisualSEEk: a fully automated content-based image query system," In *Proc. ACM Multimedia '96 Conference* (Boston, MA, USA, November 1996).

[5] VIGNA, G., Ed. *Mobile Agents and Securit*y, vol. 1419 of *Lecture Notes in Computer Sci-enc*e. Springer Verlag, Berlin Heidelberg, 1998.

[6] S.Papastavrou, E.Pitoura, and G. Samaras, "Mobile Agents for Distributed Database Access," Technical Report TR 98-12, Univ. of Cyprus, Computer Science Department, Sept. 98.

[7] Antenella Di Stefano, "Locating Mobile Agents in a Wide Distributed Environment IEEE TRNS. ON Parallel and Distributed Systems," VOL.13 pp.844-863 , Aug. 2002

[8] Dengsheng Zhang, Aylwin Wong, Maria Indrawan, and Guojun Lu Gippsland,"Content-based Image Retrieval Using Gabor Texture Features," School of Computing and Information Technology Monash University, Churchill, Victoria, 3842, Australia.

[9] M.L Kherfi and D.Ziou, and A.Bernardi, "Image Retrieval From the World Wide Web:Issues Techniques, and Systems," ACM Computing Surveys, vol.36No.1, March 2004,pp.35-67.

[10] C. J. Lee, S. D. Wang, 'Fingerprint feature extraction using Gabor filters', Electronics Letters of the IEE (UK), vol. 35 no. 4 pp. 288–290, 18 Feb. 1999

[11] F. A. P. Petitcolas, R. J. Anderson, M. G. Kuhn, 'Information hiding – a survey' Proceedings of the IEEE (USA), vol.87 no. 7 pp. 1062–1078, July 1999,

[12] Sabu.M.Thampi, Dr. K Chandra Sekaran. *Mobile Agents for Content Based WWW Distributed Image Retrieval. Proceedings of ICHMI 2004, December 2004, Bangalore.*